\documentclass[conference]{IEEEtran}

\IEEEoverridecommandlockouts
\usepackage{cite}
\usepackage{amsmath,amssymb,amsfonts}
\usepackage{algorithmic}
\usepackage{graphicx}
\usepackage{textcomp}
\usepackage{xcolor}
\usepackage{url}
\usepackage{amssymb,amsmath}
\usepackage{colortbl}

\usepackage{color,soul}
 
\usepackage{comment}
\usepackage{acronym}
\newacro{AI}[AI]{Artificial Intelligence}
\newacro{API}[API]{Application Programming Interface}
\newacro{AR}[AR]{Augmented Reality}
\newacro{BFT}[BFT]{Byzantine Fault Tolerance}
\newacro{BIIoT}[BIIoT]{Blockchain-based IoT}
\newacro{CPS}[CPS]{Cyber-Physical System}
\newacro{CPPS}[CPPS]{Cyber-Physical Production System}
\newacro{DAG}[DAG]{Directed Acyclic Graph}
\newacro{DBFT}[DBFT]{Delegated Byzantine Fault Tolerance}
\newacro{DL}[DL]{Deep Learning}
\newacro{DLT}[DLT]{Distributed Ledger Technology}
\newacro{DoF}[DoF]{Degree of Freedom}
\newacro{DoS}[DoS]{Denial of Service}
\newacro{DPoS}[DPoS]{Delegated Proof-of-Stake}
\newacro{ECC}[ECC]{Elliptic Curve Cryptography}
\newacro{ERP}[ERP]{Enterprise Resource Planning}
\newacro{GDPR}[GDPR]{General Data Protection Regulation}
\newacro{GPU}[GPU]{Graphics Processing Unit}
\newacro{GPUs}[GPUs]{Graphics Processing Units}
\newacro{HMD}[HMD]{Head-Mounted Display}
\newacro{IAR}[IAR]{Industrial Augmented Reality}  
\newacro{ICPS}[ICPS]{Industrial Cyber-Physical Systems}
\newacro{IIoT}[IIoT]{Industrial Internet of Things}
\newacro{IMU}[IMU]{Inertial Measuring Unit}
\newacro{IoT}[IoT]{Internet of Things} 
\newacro{IPFS}[IPFS]{Interplanetary File System}
\newacro{IVR}[IVR]{Industrial Virtual Reality}
\newacro{M2M}[M2M]{Machine-to-Machine}
\newacro{MES}[MES]{Manufacturing-Execution System}
\newacro{MQTT}[MQTT]{Message Queuing Telemetry Transport}
\newacro{ML}[ML]{Machine Learning}
\newacro{NLP}[NLP]{Natural Language Processing}
\newacro{PBFT}[PBFT]{Practical Byzantine Fault Tolerance}
\newacro{P2P}[P2P]{Peer-to-Peer}
\newacro{PLC}[PLC]{Power Line Communication}
\newacro{PLM}[PLM]{Product Lifecycle Management}
\newacro{PoA}[PoA]{Proof-of-Activity}
\newacro{PoB}[PoB]{Proof-of-Burn}
\newacro{PoP}[PoP]{Proof-of-Personhood}
\newacro{PoS}[PoS]{Proof-of-Stake}
\newacro{PoW}[PoW]{Proof-of-Work}
\newacro{NFT}[NFT]{Natural Feature Tracking}
\newacro{RFID}[RFID]{Radio Frequency Identification}
\newacro{RSA}[RSA]{Rivest–Shamir–Adleman}
\newacro{NSA}[NSA]{National Security Agency}
\newacro{SCM}[SCM]{Supply Chain Management}
\newacro{SCP}[SCP]{Stellar Consensus Protocol}
\newacro{SDK}[SDK]{Software Development Kit}
\newacro{SDN}[SDN]{Software Defined Networking}
\newacro{TaPoS}[TaPoS]{Transactions as Proof-of-Stake}
\newacro{TLS}[TLS]{Transport Layer Security}
\newacro{VR}[VR]{Virtual Reality}
\newacro{WSN}[WSN]{Wireless Sensor Network}

 \def\BibTeX{{\rm B\kern-.05em{\sc i\kern-.025em b}\kern-.08em
    T\kern-.1667em\lower.7ex\hbox{E}\kern-.125emX}}

\begin{document}

\title{Fake News, Disinformation, and Deepfakes: Leveraging Distributed Ledger Technologies and Blockchain to Combat Digital Deception and Counterfeit Reality \\
 
\thanks{This work has been funded by the Xunta de Galicia (ED431G2019/01), the Agencia Estatal de Investigación of Spain (TEC2016-75067-C4-1-R, RED2018-102668-T, PID2019-104958RB-C42) and ERDF funds of the EU (AEI/FEDER, UE).}
}
 

\author{\IEEEauthorblockN{Paula Fraga-Lamas, Tiago M. Fern\'andez-Caram\'es}
\IEEEauthorblockA{Dpt. of Computer Engineering, Centro de investigaci\'on CITIC, \\ Faculty of Computer Science, Universidade da Coru\~na, 15071, A Coru\~na, Spain}
Email: paula.fraga@udc.es, tiago.fernandez@udc.es}
 
\maketitle

\begin{abstract}
The rise of ubiquitous deepfakes, misinformation, disinformation and post-truth, often referred to as fake news, raises concerns over the role of Internet and social media in modern democratic societies. Due to its rapid and widespread diffusion, digital deception has not only an individual or societal cost, but it can lead to significant economic losses or to risks to national security. 
Blockchain and other Distributed Ledger Technologies (DLTs) guarantee the provenance and traceability of data by providing a transparent, immutable and verifiable record of transactions while creating a peer-to-peer secure platform for storing and exchanging information.
This overview aims to explore the potential of DLTs to combat digital deception, describing the most relevant applications and identifying their main open challenges. Moreover, some recommendations are enumerated to guide future researchers on issues that will have to be tackled to strengthen the resilience against cyber-threats on today's online media.
\end{abstract}



\section{Introduction}

Gartner predicts that the majority of individuals in developed economies will consume more false than true information by 2022 \cite{Panetta2017}. 
Digital deception is commonly recognized as deceptive or misleading content created and disseminated to cause public or personal harm (e.g., post-truth, populism, satire) or to obtain a profit (e.g., clickbaits, cloaking, ad farms, identity theft). In the context of mass media, digital deception originates either from governments or non-state actors that publish content without economic or educational entrance barriers. As a consequence, these horizontal and decentralized communications cannot be controlled or stopped with traditional centralized tools. In addition, this lack of supervision allows for security attacks (e.g., social engineering). Moreover, the veracity of information seems to be sometimes negotiable for the sake of profit, as the competition is increasingly tough. 

While trust in mass media and established institutions is declining, the use of social media is rising sharply and it has become an important source for the distribution of digital deception. Today, social media platforms miss an adequate regulation and their responsibilities are still not clearly defined. A number of issues are open \cite{EUreport}, like the application of adequate data protection rules (e.g., \ac{GDPR}) along with the market concentration in just a few social media companies worldwide.

Advances in \ac{AI} have recently been used to create sophisticated disinformation. As a result, a number of research projects as well as regulations have been launched to detect digital deception \cite{Wardle2017}. Nevertheless, researchers claim that ubiquitous content can be hardly supervised.
 
Today, Distributed Ledger Technologies (DLTs) and specifically blockchain, present challenges but also opportunities for stakeholders and policymakers as potential technologies that can help to combat digital deception. These technologies enable privacy, security and trust in a decentralized Peer-to-Peer (P2P) network without any central managing authority.  DLTs ability to combat digital deception is focused on controlling the traceability of the media, the communications architecture and the transactions.
However, the problems involved in developing effective ways to identify, test, transmit and audit information are still open.

There are only a few articles of the literature that use blockchain to combat digital deception and counterfeit reality, and they are mostly focused on tracing the source of the information. To the knowledge of the authors, this is the first article that proposes a global vision on how to confront fake news and deepfakes through DLTs with the aim of guiding researchers and managers on future developments.
Thus, this article provides a comprehensive overview on the applicability of DLTs to tackle digital deception, 
showing the potential of DLTs for revolutionizing the media industry.

The rest of this paper is organized as follows. 
Section \ref{section:stateoftheart} provides an overview of current digital deception and the involved technologies.
Section \ref{section:iniciatives} lists different DLT-based applications to combat digital deception and counterfeit reality. 
In Section \ref{section:challenges}  the main challenges of the application of DLT to tackle digital deception are analyzed and some recommendations are proposed. 
Finally, Section \ref{section:conclusions} is devoted to conclusions.



\section{State-of-the art} \label{section:stateoftheart} 
 
\subsection{Main characteristics of fake news and deepfakes}
 

Fake news is a type of disinformation (i.e., false information that is spread to deceive) that, to our best knowledge, is currently generated manually. The term was coined due to the controversies witnessed during the 2016 U.S. presidential election \cite{EUreport,Shae2019}. Nowadays, there is no consensus on the definition of fake news, since it may depend on the field of study (e.g., ethics, neuroscience, economics) or in the subjective point of view of a user. Generally, fake news can be understood as distorted signals uncorrelated with the truth. 

The term deepfakes referred originally to manipulated videos with face-swapping techniques. Nevertheless, the underlying techniques are rapidly evolving to fabricate fictional events \cite{Shae2019}.



Fake news and deepfakes in the context of digital deception are characterized by the following main elements \cite{EUreport}:

\begin{itemize}
    
    \item Type of information: it covers matters of public interest (e.g., politics, health, environment).
    
    \item Intention of the author: the content is designed to wholly or partially deceive, manipulate or mislead, or it utilizes unethical persuasion techniques (e.g., propaganda or ideology-driven content).

    \item Dissemination strategy: its information is disseminated strategically through automated and aggressive techniques (e.g., campaign-like manners, fake accounts, bots, micro-targeting or trolls). 
    
    \item Consequences of dissemination: the spread is focused on generating insecurity, hostility or polarization, or it attempts to disrupt democratic processes (e.g., elections, referendums), fundamental rights or the rule of law. Nevertheless, although the potential influence and impact of fake news and deepfakes still remains uncertain, in at least a few cases (e.g., the Brexit campaign, the independence of Catalonia), they appear to have impacted significantly public behavior \cite{EUreport}. Another example is the so-called misinfodemics, where health misinformation (e.g., the effect of vaccines, the outbreak of coronavirus COVID-19) may enable the spread of diseases.  



    \item Inner characteristics: it has characteristics that enable rapid and widespread diffusion. In fact, fake news and deepfakes are likely to spread faster and further than the truth \cite{Vosoughi2018}. It is increasingly resistant to detection as enablers such as AI, \ac{IoT}, \ac{AR} and \ac{VR} are progressing rapidly. 
\end{itemize}

 

\subsection{The role of emerging technologies}

Forging information has never been easier thanks to a range of free content-generation software.
Moreover, with emerging technologies like \ac{IoT}, people is more exposed to being monitored. Furthermore, the capacity of \ac{AR} and \ac{VR} to imitate reality is growing and can be harmful, since immersive experiences are less subject to rational thinking and they amplify the effects of potential manipulations.

In addition, \ac{NLP} and AI are expected to drive the upcoming counterfeit reality, where detecting manipulation will become almost impossible for people and more complicated for machines \cite{Kim2018}. For instance, \ac{DL} is being increasingly used to create models such as Generative Adversarial Networks (GANs) that enable realistic manipulations of image and video that are unrecognizable by both human and machines. 

Furthermore, the emergence of deepfakes will exacerbate significantly the impact of digital deception. Individuals, businesses and society as a whole may face novel forms of extortion as well as additional risks for democracy and national security.

\subsection{The role of new media}
Social media organize and amplify the effect of content communication.
Citizens may believe that the content they consume is user-generated, spontaneous, neutral and universal, while the truth is that such a content may have been provided strategically and micro-targeted \cite{EUreport}.
In addition, social media privacy policies and terms of service allows for collecting citizens' big data  (e.g., patterns, profiles) to sell them to a number of actors for massive profiling, advanced demographic analytics, micro-targeted advertising and the automation of content. For instance, lack of transparency hinders advertiser traceability (as they may deliberately hide their identity or use intermediaries) and makes it more difficult to obtain digital evidence to reinforce liability.

\begin{figure*}[!htb]
\centering
\includegraphics[scale=0.3]{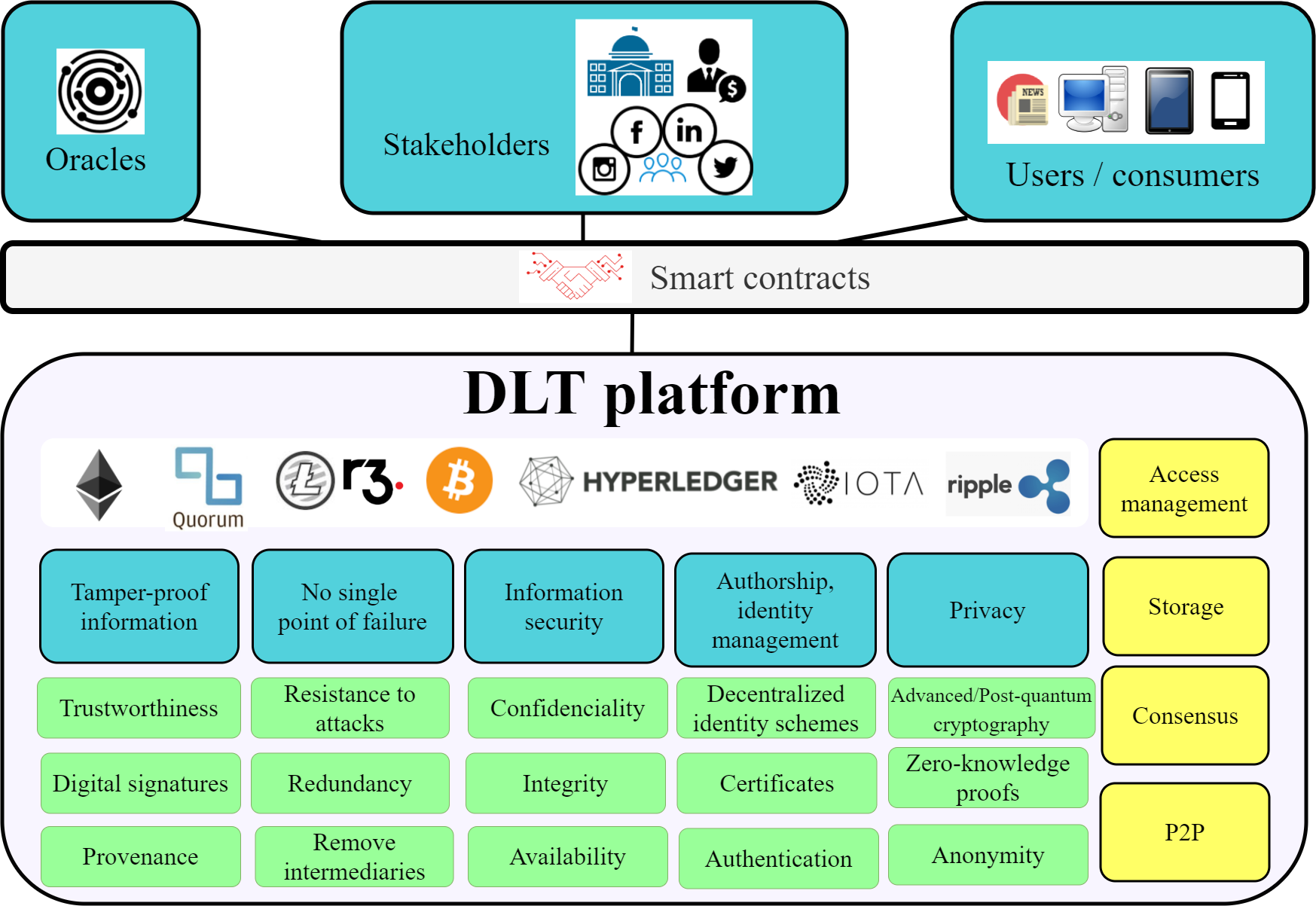}
\caption{DLT and blockchain key capabilities to combat digital deception.}
\label{fig:BCcapabilities}
\end{figure*}

\subsection{DLTs and blockchain capabilities}  


DLTs like Tangle or blockchain are able to provide seamless authentication, efficient and secure data storage, processing and sharing, robustness against attacks, scalability, transparency and accountability. 
Such features (illustrated in Figure \ref{fig:BCcapabilities}), together with the use of smart contracts enabled by oracles, can play an effective role in combating fake news and deepfakes, considering that transactions cannot be tampered once they have been distributed, accepted and validated by a network consensus  \cite{Shahaab2019} and stored in blocks. Moreover, transactions are easily auditable by all the involved stakeholders.

\section{DLT-based applications to combat digital deception}  \label{section:iniciatives}  


There are just a few articles in the literature that study the applicatibility of DLTs to face fake news and deepfakes, all of them are quite preliminary and just focus only on a specific application. In this section we provide an overview of the most promising solutions to leverage DLTs to identify, prevent and detect digital deception:

 \begin{itemize}
\item{Decentralized} content moderation:
conventional content moderation processes (e.g., flagging, notice and take down) rely on a centralized regulator with immediate content removal capabilities.
In DLTs, especially in the case of permissionless ledgers, anyone can participate or become a transaction validator and there is no central authority, therefore additional consensus mechanisms should be implemented.


\item{Trustworthiness checkers:}
Qayyum et al. \cite{Qayyum2019} introduced the concept of Proof-of-Truthfulness (PoT), where any node in the network can verify whether a content is or not part of a blockchain. Content is stored in a Merkle tree, a binary tree built using hash pointers in which nodes at the $n-1$ level contain hash pointers to the content stored at the $n$ level. Given a specific content, its trustworthiness could be verified in $O(log(n))$ by searching throughout a single tree branch from the content to the root (level $0$).

\item{Fact-checking incentivized dApps:}
reliable fact-checkers \cite{Zhang2020} can be identified (since they are interested in validating content) so they can get financial rewards (e.g., tokens), as well as to increase their reputation for high-quality work. The amount of received rewards increases as the fact-checker improves his/her/its reputation. In such a system, content creators will be also interested in submitting their content for validation in order to build their reputation.

\item{Reputation systems:}
a score can be used for measuring the credibility of a publisher and warn readers when the content shows traits that may indicate biases. In \cite{Qayyum2019}, it is proposed a dynamic reputation set: an initial zero score is assigned to each non-verified media and the score evolves as the entity shares trustworthy verified news.
Registered consumers provide feedback through the platform or score the credibility of the content, like in the case of BitPress \cite{bitpress}. Nevertheless, the problem of subjectivity, bias and the risk of malicious actors have to be further studied.

\item{Community-driven dApps:}
crowdfunding approaches can use tokens to incentivize the discovery of truth.
In DLT-based social networks, users can exchange tokens or coins through the same social network in a straightforward way.
For instance, users can perform secure P2P transactions without third-party intermediaries through cryptographically signed smart contracts.

\item{Decentralized social media platforms:}
the Solid project \cite{solid} proposes a set of tools for building social decentralized applications (dApps) based on Linked Data principles, resulting in improved privacy as well as true data ownership, access control and storage location.

Another interesting initiative is the Content Blockchain Project \cite{iRights}, an open and decentralized blockchain ecosystem for the distribution of media content operated and owned by the industry itself. The main element of the project is the International Standard Content Code (ISCC), which is similar to identifiers like the International Standard Book Number (ISBN), but with enhanced functionality in order to create a user-friendly application that generates ISCCs without any cost. 

A social media platform may also be re-engineered as a  Decentralized Autonomous Organization (DAO). DAOs enable self-organization and self-governance by encoding operational and managerial rules on a blockchain through smart contracts. However, DAOs still face many challenges, such as security and privacy issues and an unclear legal status \cite{Wang2019}.
A smart contract can add functionality to a DLT as it is a computer program that is stored in the distributed database \cite{Hasan2019}. Smart contracts allow for the addition of validations, constraints and business logic to transactions in a form of an agreement between parties. 
Moreover, smart contracts can be used to register, update and revoke the identities of different organizations (e.g., publishers), as well as to determine their status and reputation score \cite{Qayyum2019}.

\item{Additional platform-based services:} 
online platforms (e.g., Mozilla, Facebook, Twitter) and trade associations (e.g., EACA, IAB Europe, WFA) have made some progress in their commitment to tackle fake news \cite{ECreports}. For instance, Google has announced the Google News Initiative to support the news industry in quality journalism \cite{Shae2019}.

\item{Notarization services:}
automatic management of non-tampered content and multi-node content verification can help to overcome the problem of verifying big data news streams.
DLTs inherently guarantee data integrity once transactions are stored. This feature makes DLTs an essential infrastructure for notarization services \cite{Song2019}. Nevertheless, a central problem is how to ensure that data are not forged before they are added into a block. 

\item{Provenance and ownership services:}
the use of DLT technology would also allow for making content forgery almost impossible by demonstrating its origin and, in case of detecting a counterfeit, it would make the owner accountable. For instance, Huckle et al. \cite{Huckle2017} proposed an Ethereum framework with standardized metadata for the verification of the authenticity and the provenance of digital media. 
However, the ability of the proposed system to find fake resources is somewhat limited (i.e., it is not able to prove the authenticity of a story as a whole).
 A similar solution is described in \cite{Huckle2017}, where the authors present an early prototype of the `Provenator', which stores provenance metadata (i.e., objects, events, agents and rights) on the Ethereum blockchain and allows users to check the provenance of media resources. 

\item{Traceability and tracking services:}
Shang et al. \cite{Shang2018} trace the source of news by keeping a ledger of timestamps and the links between the different blocks.
The proposed procedure is as follows. First, when media are writing news, the related content, category and other information are uploaded to the blockchain. 
Then, in the process of communicating the news, it is recorded the release date, the hash value and the timestamp of the pre-block so that the chain structure can be formed. Third, when readers consume news, they can trace the source through the chain structure of the blockchain and the stored information. 
Although this scheme seems promising, the authors point out that the construction of the whole traceability system needs to be further explored.
 
Another relevant proposal can be found in \cite{Hasan2019}, where the authors present an Ethereum-based solution for multimedia history tracking that uses \ac{IPFS}, an Ethereum name service, a reputation system and smart contracts.

\item{Forensics:}
it is challenging to make sure that devices, content and intellectual property are legitimately used with authorization and to prove forensically with a certain degree of confidence when otherwise. Once security is compromised, if there is an intellectual right infringement or a counterfeit, forensics can recreate what has happened to answer what, when, who, where, and how. DLT-based notary services offer unarguable digital evidence because the integrity of the content has been cryptographically guaranteed.

\end{itemize}

As a conclusion, it can be stated that this section described a broad range of DLT-based applications that can be used independently or combined, each with different technical requirements in terms of robustness, scalability, performance, interoperability, or privacy. In addition, note that the proposed applications can deal with all types of media content, but the majority of the cited academic solutions were designed to target fake news (i.e., text). Furthermore, it is worth mentioning that, in the short-term, the greatest impact will come from traceability and tracking services implemented by start-ups or additional platform-based services from big media platforms. Nevertheless, more disruptive solutions like decentralized social media platforms cannot be neglected.

 

\section{Challenges and recommendations}  \label{section:challenges}

The following are the most relevant open challenges and recommendations for guiding future researchers, developers and managers to combat digital deception:

\begin{itemize}
    \item The current efforts of the research community are mostly focused on one type of fake news (i.e., verifiable false content), while other bad practices are barely studied.

    \item Most digital deception detection proposals are based on cryptographic hashes, which are sensitive to noise and, when there is a change of a character, a pixel or a bit in a certain content, it can result in a different hash \cite{Huckle2017}. While any minimal change in two resources will generate vastly different hashes, the use of perceptual hashes produces comparable results if the resources are similar. 
    Another alternative to overcome this problem is the use of a semantic similarity index of a content published by different sources. 

    \item 
    The DLT design must be optimized for the specific use case, which should consider the level of required decentralization and the consensus algorithms since they will impact performance (e.g., transaction throughput) and scalability.

     \item Strengthening cybersecurity and preserving privacy and security of content shared on social media is also a key issue, since it may be used to train an ML/DL model to create fake content. DLT-based solutions can cryptographically store the content in such a way that every transaction and interaction with it is traceable.  
     
     \item  Most of the current cryptography used by DLTs is vulnerable to certain quantum computing attacks, so post-quantum blockchain solutions must be further investigated \cite{PQBC}.
 
    \item There are still open issues related to the DLT compliance with GDPR \cite{GDPR}, especially  when dealing with the role of the controller, the feasibility of data anonymization and the ease of subject rights. 

    \item Future platforms will have to ensure safety and transparency by providing a trade-off between content moderation (e.g., freedom of expression, right to receive information) and personal data protection.  In addition, there are still concerns on the fact that social interactions and transactions may be mediated by trustless technological systems controlled by a few dominant players.

    \item The identification of digital deception and counterfeit reality is a rapidly evolving challenge that requires multidisciplinary collaborations (e.g., industry, governments and media). Moreover, there is no one size fits all solution for the general intervention mechanisms (e.g., personalized solutions).
    
     \item A DLT-based system alone is not able to fully evaluate the authenticity of a content. Consequently, it is essential to develop a system that is resilient to data falsification attacks, which inserts forged data into the DLT. To face this issue, it is recommended to include contextual knowledge to corroborate the integrity of the media (e.g., social context features, domain location, temporal patterns). Further research may include the use of DLT together with AI and NLP methods to develop deep insights about similarities and to quantify trustworthiness.

   \item AI’s ability to detect digital deception in media is lower than its ability to create it \cite{Shae2019}. The accuracy of current techniques heavily depends on the training datasets and the underlying algorithms and protocols.
   Furthermore, social media platforms, which make use of complex interactions and information flows, require a variety of DLT-based trust mechanisms and novel AI techniques to prevent deception \cite{Shae2019}. The long-term aim should be to devote strategies to prevent counterfeit reality before its spreading.

\end{itemize}


\section{Conclusions} \label{section:conclusions}
Provenance, consensus and traceability can be guaranteed with DLTs when creating a P2P platform for tackling digital deception. This article analyzed some applications currently under development and proposed a number of additional mechanisms to control content. Although there are technological and practical limitations of DLT technology when combating digital deception, the trust mechanisms provided by DLT can make it more adequate than other technologies for ensuring authenticity and auditing, enabling accountability and eliminating counterfeit reality. Moreover, future researchers are encouraged to develop joint AI and DLT solutions in an enhanced coordinated effort to address all the aspects of digital deception.

\begin{IEEEbiography}[{\includegraphics[width=1in,height=1.25in,clip,keepaspectratio]{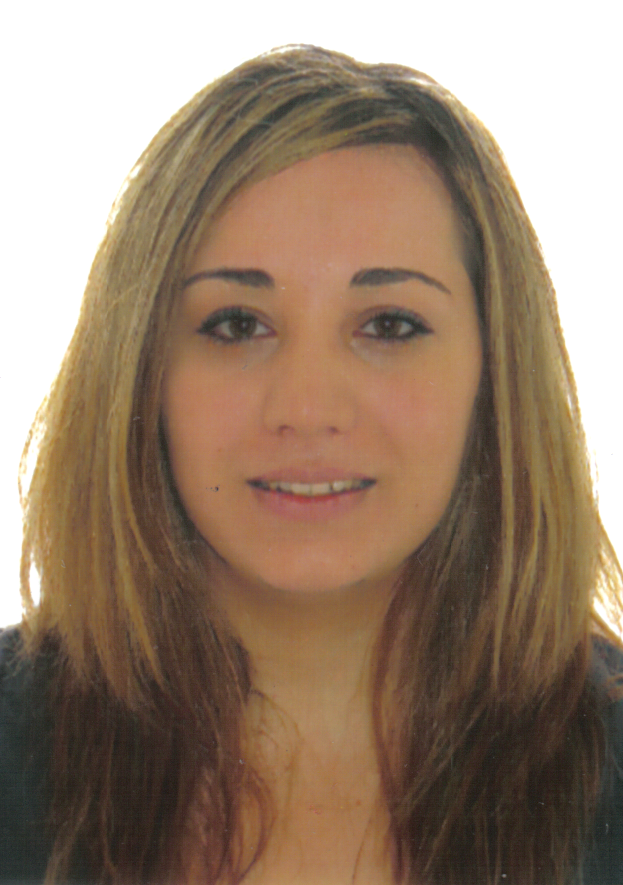}}]{Paula Fraga-Lamas}(M'17) received the M.Sc. degree in Computer Engineering from the University of A Coru\~na (UDC) in 2009, and the M.Sc. and Ph.D. degrees in the joint program Mobile Network Information and Communication Technologies from 5 Spanish universities: University of the Basque Country, University of Cantabria, University of Zaragoza, University of Oviedo and University of A Coru\~na,  in  2011 and 2017, respectively. She holds an MBA and postgraduate studies in business innovation management (Jean Monnet Chair in European Industrial Economics, UDC), Corporate Social Responsibility (CSR) and social innovation (Inditex-UDC Chair of Sustainability). Since 2009, she has been with the Group of Electronic Technology and Communications (GTEC), Department of Computer Engineering (UDC). She has over 60 contributions in indexed international journals, conferences, and book chapters, and 4 patents. She has also been participating in over 30 research projects funded by the regional and national government as well as R\&D contracts with private companies. She is actively involved in many professional and editorial activities, acting as reviewer, advisory board member, topic/guest editor of top-rank journals and TPC member of international conferences.
Her current research interests include mission-critical scenarios, Industry 4.0, Internet of Things (IoT), Cyber-Physical Systems (CPS), Augmented Reality (AR), fog and edge computing, blockchain and Distributed Ledger Technology (DLT), and cybersecurity.
\end{IEEEbiography}

\begin{IEEEbiography}[{\includegraphics[width=1in,height=1.25in,clip,keepaspectratio]{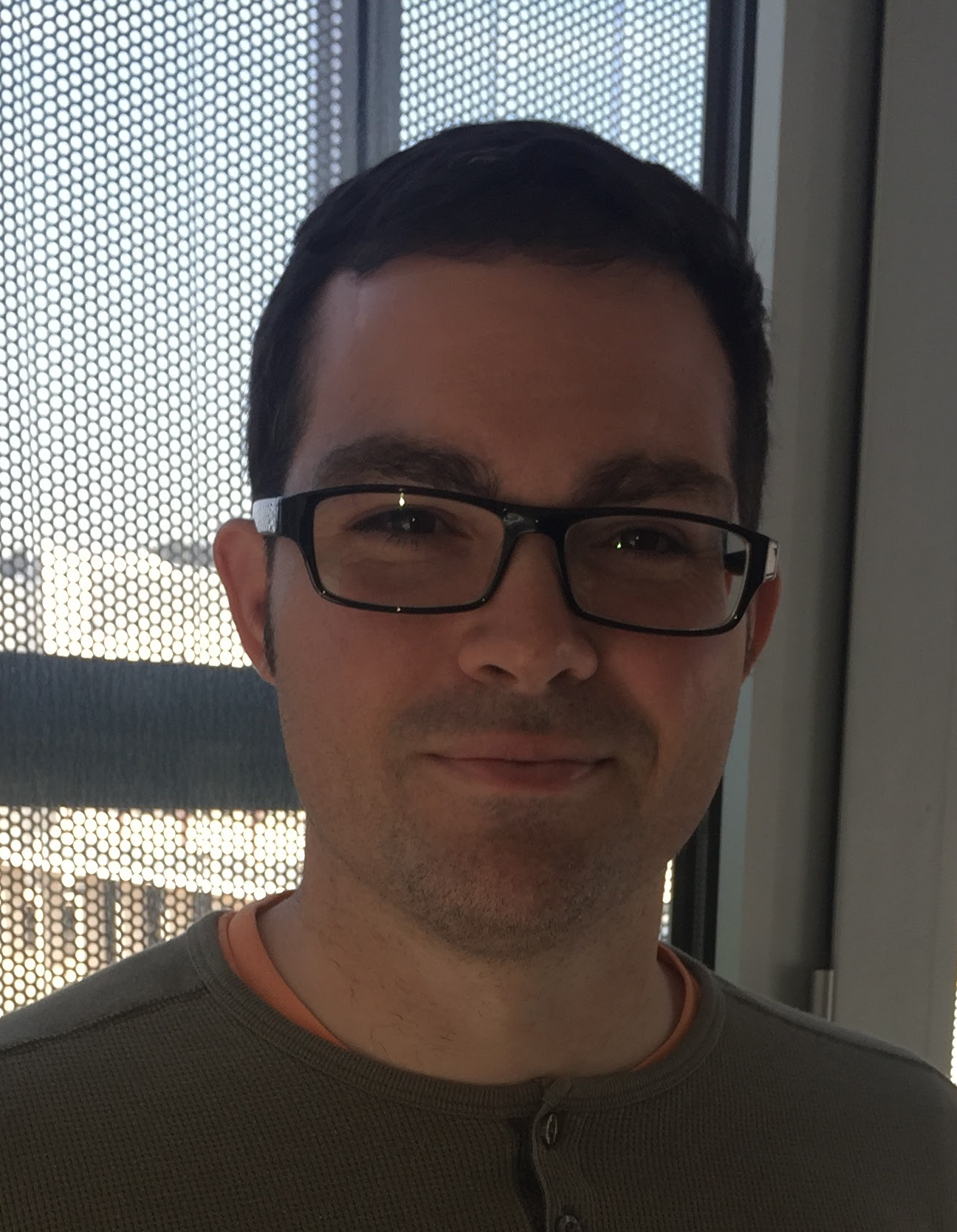}}]{Tiago M. Fern\'andez-Caram\'es} (S'08-M'12-SM'15) works since 2016 as an Assistant Professor in the area of Electronic Technology at the University of A Coruña (UDC) (Spain), where he obtained his MSc degree and PhD degrees in Computer Science in 2005 and 2011. Since 2005 he has worked in the Department of Computer Engineering at UDC: from 2005 to 2009 through different predoctoral scholarships and between 2007 and 2016 as Interim Professor.
His current research interests include IoT/IIoT systems, RFID, wireless sensor networks, augmented reality, embedded systems and blockchain, as well as the different technologies involved in the Industry 4.0 paradigm. In such fields, he has contributed to 40 papers for conferences, to 35 articles for JCR-indexed journals and to two book chapters. Due to his expertise in the previously mentioned fields, he has acted as peer reviewer and guest editor for different top-rank journals, and as project reviewer for national research bodies from Austria (FWF), Croatia (CSF) and Argentina (ANCyT). 
\end{IEEEbiography}


\end{document}